\documentclass[aps,prl,twocolumn,superscriptaddress]{revtex4}
\usepackage{amsfonts}
\usepackage{amsmath}
\usepackage{amssymb}
\usepackage{graphicx}
\usepackage{float}

\usepackage{color}
\begin{document}
\title{Simulations of supercooled water under  passive or active stimuli} 


\author{Victor Teboul}
\email{victor.teboul@univ-angers.fr}
\affiliation{Laboratoire de Photonique d'Angers EA 4464, Universit\' e d'Angers, Physics Department,  2 Bd Lavoisier, 49045 Angers, France.\\
*Electronic mail: victor.teboul@univ-angers.fr}

\author{Gabriel Rajonson}
\affiliation{Laboratoire de Photonique d'Angers EA 4464, Universit\' e d'Angers, Physics Department,  2 Bd Lavoisier, 49045 Angers, France.\\
*Electronic mail: victor.teboul@univ-angers.fr}


\begin{abstract}

We use molecular dynamics simulations to study the behavior of supercooled water subject to different stimuli from a diluted azobenzene hydrophobic probe. When the molecular motor doesnÕt fold, it acts as a passive probe, modifying the structure of water around it, while when the motor is active, it induces elementary diffusion processes inside the medium acting mainly on the dynamics. We study two particular densities, the density of ambient water and a lower density around the ambient pressure ice density, chosen to favor HDL and LDL water respectively. We find that the passive probe induces ever an acceleration or a slowing down of the diffusion process around it depending on the density of water, while the active probe induces acceleration only. We find a crossover between the diffusion coefficients for the two densities near the passive probe, around $T = 215K$. This dynamical crossover is associated to a modification of the structure of water near the probe. Structure calculations show a crossover of the proportion of LDL water around the same temperature suggesting that it induces the observed dynamical crossover. In opposition with these results, the active stimuli increase diffusion for both densities and decrease the proportion of LDL water at low temperature. However we also find for the active stimuli a crossover of the LDL proportion between the two densities of study, showing remarkable similarities between active and passive stimuli results.

\end{abstract}

\maketitle
\section{ Introduction}

While of quite simple structure and composition, water is a strange liquid\cite{anomalies2}, displaying a number of anomalies\cite{anomalies,anomalies2}.
Water is also important as the liquid where biological processes take place.
The large number of anomalies relates to the presence of different possible structures (a property known as polyamorphism), and also to 
  hydrogen bonding that leads to the formation of a network organized liquid. 
Three decades ago, to explain water anomalies, Poole et al. \cite{LDL2} postulated the existence of two different liquid states of water, a low density liquid (LDL)  and a high density liquid (HDL), with a critical point located at low temperature in the supercooled region \cite{LDL0,LDL1,LDL2,LDL3,LDL4,LDL5,LDL6,LDL7,LDL8,LDL9}. 
Because below the homogeneous nucleation temperature $T_{H}=232 K$ water crystallizes rapidly \cite{valeria1,valeria2,valeria3,nucleus1,nucleus2}, it is quite difficult to access experimentally the region of temperature below $T_{H}$ and above the glass-transition $T_{g}=136 K$. Due to that difficulty, that region of temperature ($T_{H}>T>T_{g}$) has been called the 'no man's land'.
As the postulated liquid-liquid transition between HDL and LDL is located inside the 'no man's land' it is difficult to observe experimentally and  the metastability of these two liquid states is still the subject of controversy \cite{David1,David2}.
However there is no doubt that the presence of different competing structures in liquid water influences largely  its properties.
If cooled fast enough, liquids can avoid crystallization and remain in the liquid state below their melting temperature\cite{book0,DH}. 
If the temperature is further decreased, the viscosity of the liquid  increases rapidly and eventually, at low enough temperature the medium becomes so viscous that it behaves as a solid and  is called a glass.
The reason for that increase in viscosity  is however still unknown\cite{anderson} and the object of active researches.
As discussed above, water has two important particularities:  The presence of hydrogen bounding leading to a network structure and the polyamorphism or equivalently the existence of several metastable structures.

Due to the presence of different competing structures and the associated fluctuations we expect supercooled specific properties to be larger in water than in most other liquids, leading to a particular interest in the study of that liquid\cite{waterg,conf1,conf2,conf3,conf4,conf5,conf6,conf7,confine,bulk,mixture,pressure1,pressure2,pressure3,structure,mm}. The possible presence of a liquid-liquid transition at low temperature between HDL (high density liquid water) and LDL (low density liquid water), that can be modified by the presence of impurities, however complicates the interpretations but adds interest in the study of water, in particular on the evolution of correlation lengths with temperature. {\color{black}  We study in this paper the effect of a hydrophobic molecular motor, the disperse red one DR1 molecule, in supercooled water.
Despite the small solubility of hydrophobic molecules in water, hydrophobic molecules dissolved in water have been the subject of a number of experimental and theoretical studies\cite{d10,d11,d12,d13,d14,d15,d16,d17}. 
From the experimental point of view  the motor can be attached to a hydrophilic molecule to increase its solubility, or be attached to the wall of a pore. For applicative purpose $DR1$ molecules often appear in water as part of drug delivery complexes containing hydrophilic groups.   
More interestingly, the motor's hydrophobicity can also be used to connect the motor to an interface, that can be the wall of a hydrophobic  nanopore or the surface of water, in order to study its motion on the interface during the successive isomerizations.

}

Impurities that we call here indifferently passive stimuli or probes, and active stimuli, will induce perturbations in the structure and dynamics leading to an enrichment of the properties and acting as possible probes of the underlying physical mechanisms. 
{\color{black} Molecular dynamics and Monte Carlo simulations\cite{md1,md2,md3} are invaluable tools, together with model systems\cite{ms1,ms2,ms3,ms4,ms5} to increase our understanding of unsolved problems in condensed matter physics.
}
In this paper we use molecular dynamics simulations\cite{md1,md4} to study the behavior of supercooled water subject to various stimuli.

\section{Calculation}

A variety of intermolecular potentials exist to model water\cite{SPC,SPCE,TIP5Pa,TIP5Pb,TIP5PE,POT1,POT2} sometimes with coarse graining\cite{book2,CG1} for applications in biology. 
In this paper we model the water molecular interactions with the TIP5PE potential\cite{TIP5PE} and the long range electrostatic interaction with the Reaction field method using a cutoff radius $R_{c}=9$\AA\ and an infinite dielectric constant for distances $r>R_{c}$.
The water molecule is modeled as a rigid body and we will focus our attention on  the center of masses behavior (that is also approximately the oxygen atoms behavior as the differences in the position of the center of masses and of the oxygen atom is quite small in water).

{\color{black} We  solve the equations of motion using the Gear 4 algorithm with a time step $\Delta t=10^{-15}s$. Due to the release of energy from the motor, simulations where the motor is active are out of equilibrium.
We evacuate the energy created by the motor's folding, from the system with a Berendsen thermostat. 
We use the NVT canonic thermodynamic ensemble as approximated by that simple thermostat  for the whole set of simulations (see ref.\cite{finite2} for an evaluation of the effect of the thermostat on our calculations). Our molecular dynamics simulation program derived from Allen and Tildesley library\cite{md1}  has been tested in a number of works.
}

\begin{figure}[H]
\centering
\includegraphics[height=6.2 cm]{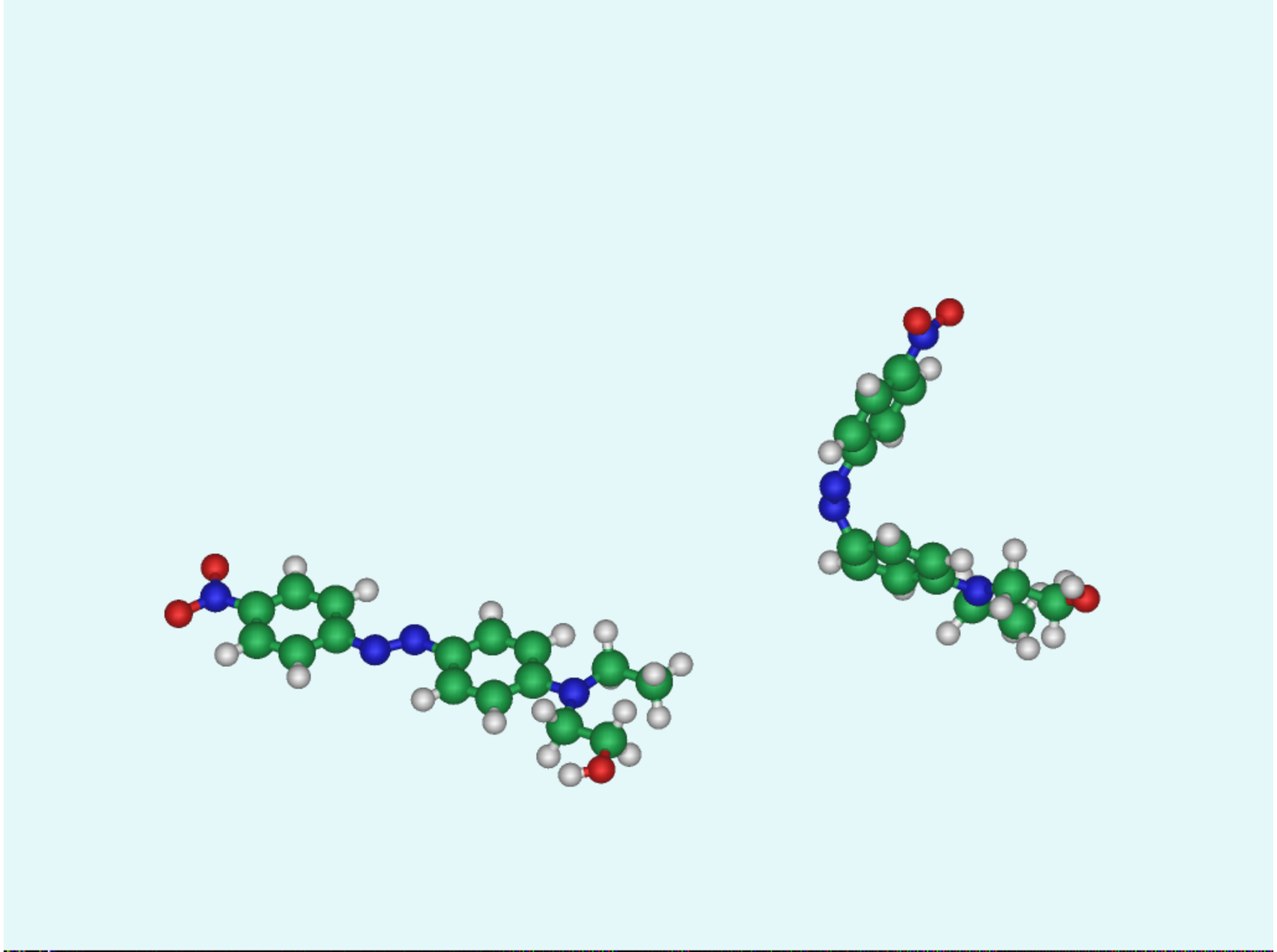}
\caption{(color online) Picture of the $DR1$ probe in the trans (left) and cis (right) forms.} 
\label{f0}
\end{figure}

\begin{figure}[H]
\centering
\includegraphics[height=6.2 cm]{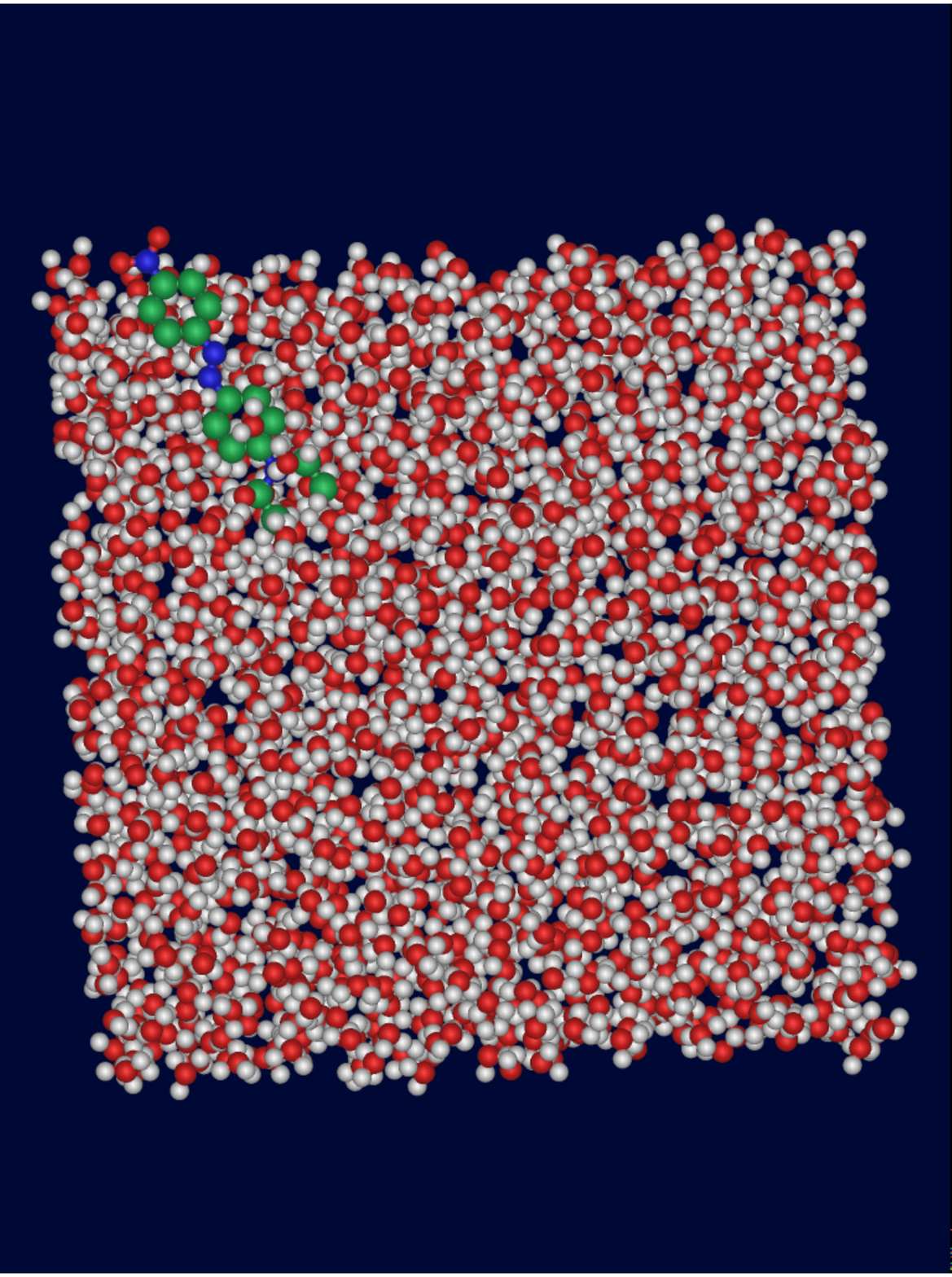}
\caption{\color{black} (color online) Picture of the water simulation box containing the $DR1$ probe. $\rho=0.925 g/cm^{3}$ and $T=210K$. } 
\label{f0b}
\end{figure}

Our simulation box of pure water contains $2000$ water molecules in a cubic box with usual periodic conditions and is aged at the temperature of study during $10ns$ (for $T \geq 250 K$) or $20 ns$ (for $T \leq 240K$) before any recorded run.  
We fix the density to $\rho=1 g/cm^{3}$ or  $\rho=0.92 g/cm^{3}$ for pure water. For the simulations using the probe we corrected the smaller density due to the presence of the probe to $\rho=0.925 g/cm^{3}$.
These densities lead to a box size of $39.10$ or $40.21$ \AA\ respectively for pure water ($39.22$ and $40.25$ \AA\ with the probe), chosen large enough to minimize finite size effects\cite{finite0,finite1,finite2} that appear at low temperatures in supercooled liquids due to the increase of  correlation length.
For the simulations of water doped with a molecular motor or probe, we add to the $2000$ water molecules, a "disperse red one" $DR1$ molecule ($C_{16}H_{18}N_{4}O_{3}$, the motor, see Figure \ref{f0}, {\color{black} and Figure \ref{f0b} for the simulation box containing the motor}) that has the property to isomerize when subject to an appropriate light stimulus. 

Notice that when diluted inside a soft medium the isomerization of the DR1 molecule was found to trigger macroscopic motions inside the medium\cite{soft,azo1,azo2,azo3,azo4,azo5,azo6,azo7,azo8,azo9}.
The origin of these motions is still a matter of controversy\cite{azo1,azo2,azo3,azo4,azo5,azo6,azo7,azo8,azo9}, but a  connection with the glass-transition problem\cite{anderson} has been suggested\cite{pre} as  induced cage-breaking processes\cite{cage} and induced dynamic heterogeneity\cite{ivt3,prl} were  reported.
The interactions of the water molecules with the probe are hydrophobic in our simulations.

For the probe molecule interaction with the medium, we use the same interaction potentials\cite{pot1} than in ref.\cite{cage,ivt4}, together with the following mixing rules
\cite{mix1,mix2}: 

\begin{equation}
\epsilon_{ij}=(\epsilon_{ii} . \epsilon_{jj})^{0.5}  ; 									
\sigma_{ij}=(\sigma_{ii} . \sigma_{jj})^{0.5}   \label{e2}
\end{equation}

When the motor is active, we model the isomerization (folding) of the probe molecule as a periodic continuous change of shape\cite{prl,cage,ivt3,ivt4} during a characteristic time $t_{0}=0.3 ps$ to the experimental cis or trans atomic positions using the quaternion formalism. The time between two foldings (approximately the half period) is chosen large enough to stay in the linear response regime.
The period of the isomerization $cis-trans$ and then $trans-cis$ is also fixed in the study $\tau_{p}=480 ps$.
During the isomerization the shape of the molecule is modified slightly at each time step using the quaternion method with constant quaternion variations, calculated to be in the final configuration after a $0.3 ps$ isomerization. This method corresponds to opposite continuous rotations of the two parts of the molecule that are separated by the nitrogen bounding. {\color{black} Experimentally, the photo-isomerization between the cis and trans isomers occurs via two possible mechanisms, the rotation and the inversion mechanisms. In our simulations we use the rotation mechanism only. However previous work\cite{x} has shown that the effect of the isomerization on the surrounding medium was only weakly dependent on the shape and mechanism of the folding. This is because the effect on the medium, is determined by a folding induced cage breaking process, a process that mainly depends on the size of the  motor relative to the medium molecules and not on the particular folding mechanism.
}
We model the isomerization to take place at periodic intervals whatever the surrounding local viscosity.
This approximation has recently been validated experimentally\cite{diamond} as the pressure that is necessary to stop the azobenze isomerization is very large ($P> 1$  $GPa$).
{\color{black} After aging runs we launch further  runs of $40$ ns each which configurations are recorded for the statistical calculations. The statistical data correspond to averages taking into account $2000$ water molecules and  $20000$ time configurations, leading to quite precise results, further validated by different runs.}

Due to the time-symmetrical folding and unfolding process  Purcell's theorem\cite{scallop1} prevents the motor's  displacement, however we are only interested here in the stimuli induced by the foldings on the liquid. Notice however that various violations of the theorem have been reported\cite{scallop2,scallop4,scallop5,scallop6,scallop7,scallop8,scallop9,scallop11,scallop12,scallop13,scallop13b,scallop14,scallop15,pre}.   Another source of motor's displacement comes from folding-induced diffusion of the medium that in turn carries the motor\cite{prl,scallop3,cage,ivt4,x,scallop10}.
We expect passive stimuli to modify the structure locally and the active induced cage-breaking stimuli to create non-thermal excitations in the medium.

\section{Results and discussion}

We will study the dynamical behavior of supercooled water doped with a molecular motor, separately when the motor is off and acts as a passive probe and when it is on and acts as an active stimulus.
Even when the motor is passive, it modifies the local structure of the medium and the dynamical local properties due to its larger mass.
Thus {\color{black} at first sight} we expect a slow down of supercooled water around the passive motor, because due to its mass its mean velocity is smaller than the mean water molecules, and it thus acts as an obstacle to the motion of water molecules. However we will see that this expectation is not always correct, due to the structural modification that sometimes acts in the opposite way. 
When the motor is active, it creates non-thermal excitations in the medium, that permit to access low temperatures while still being ergodic.

\subsection{Passive stimulus}
\subsubsection{Dynamical effects}

\begin{figure}
\centering
\includegraphics[height=7.0 cm]{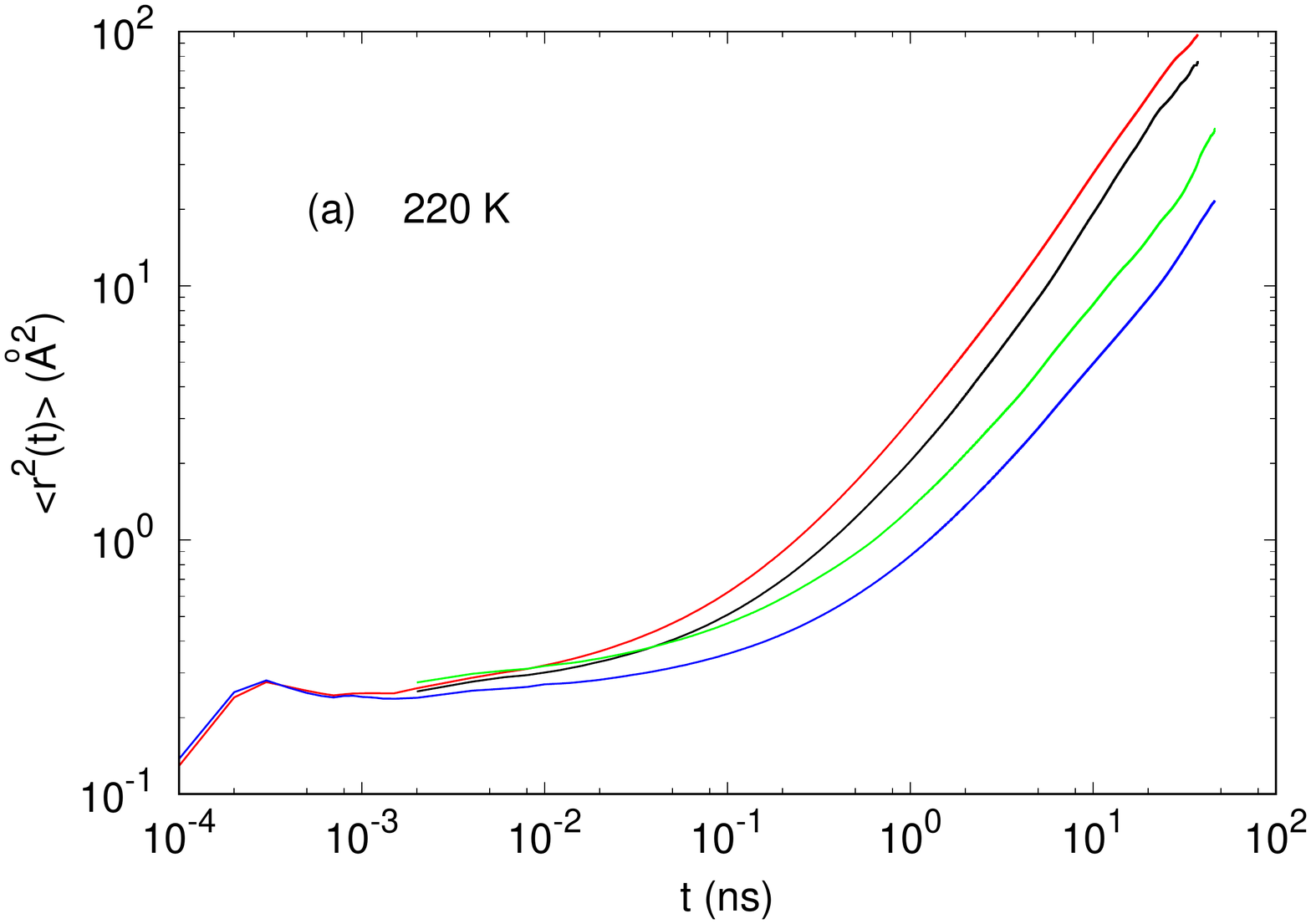}
\includegraphics[height=7.0 cm]{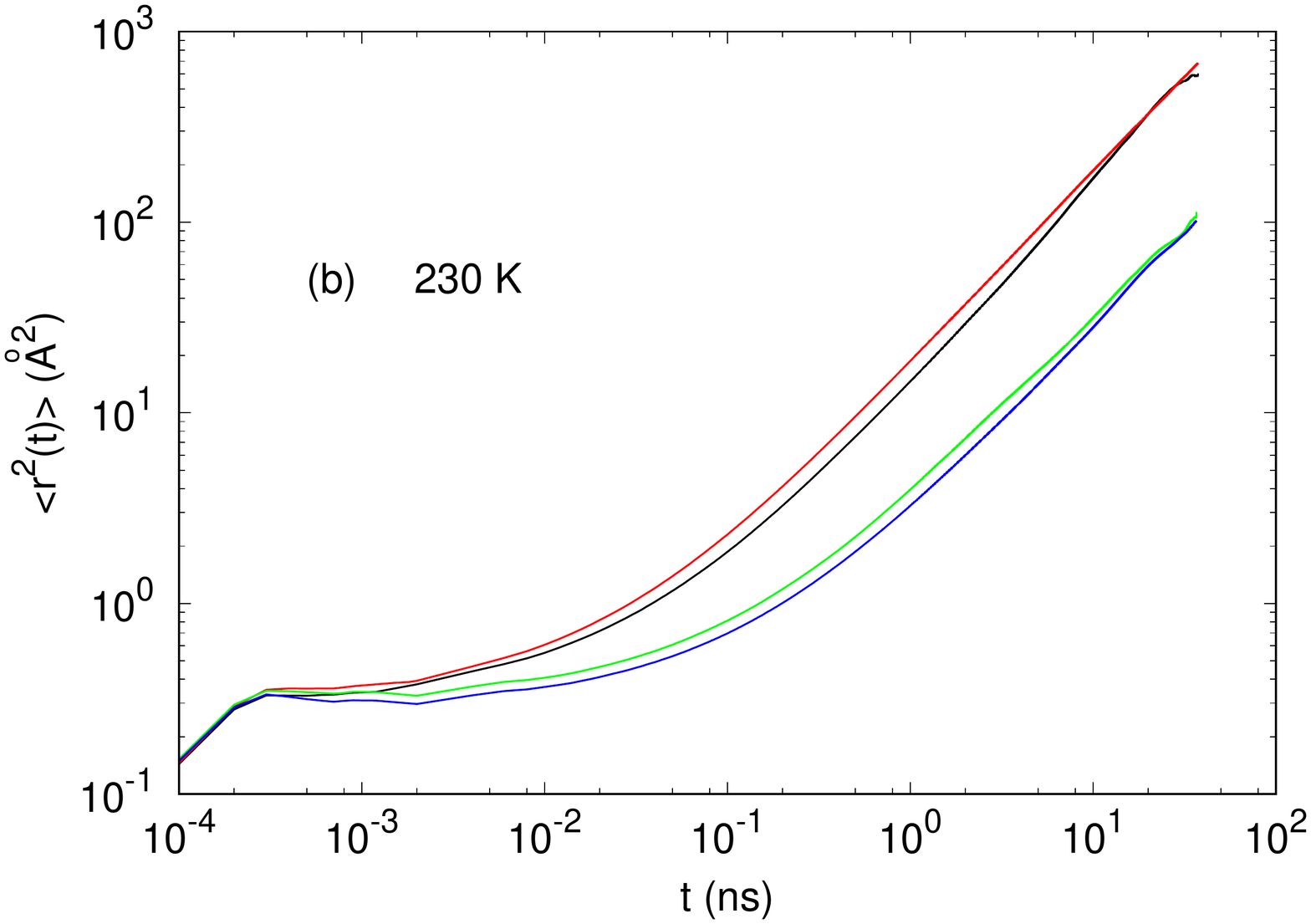}
\includegraphics[height=7.0 cm]{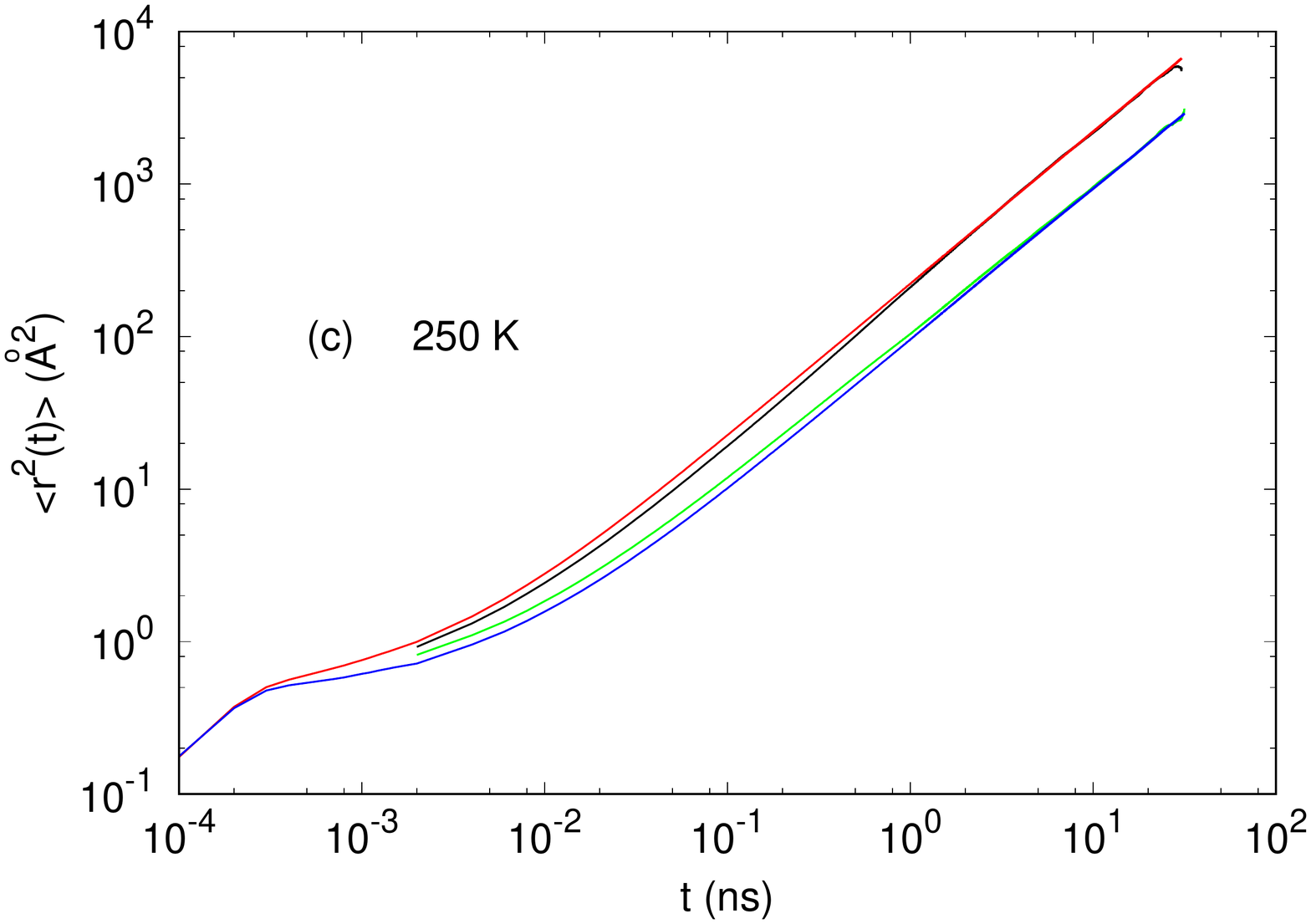}
\caption{(color online) Mean square displacement (MSD) of the center of masses of water molecules, for various temperatures.
Around the motor the displacement increases or decreases depending on the density.
(a) $T=220 K$, (b) $T=230 K$, (c)  $T=250 K$.
From top to bottom in each Figure: 
Red line: bulk water MSD for $\rho=1 g/cm^{3}$. 
Black line: water MSD around the probe ($r<10$\AA\ at $t=0$) for $\rho=1 g/cm^{3}$. 
Green line: water MSD around the probe ($r<10$\AA\ at $t=0$) for $\rho=0.925 g/cm^{3}$. 
and Blue line: bulk water MSD for $\rho=0.925 g/cm^{3}$. 
The probe effect increases at low temperatures. The probe effect is also larger at low density for the lower temperature displayed.} 
\label{f2}
\end{figure}

When the motor is passive and acts as a probe, we {\color{black} usually} expect a slow down of the dynamics.
However Figure \ref{f2} shows that depending on the water's density we observe at low temperatures an acceleration or a slowing down around the probe. 
The difference between the mean square displacements appears at low temperatures during the plateau time regime.
That result shows that water molecules escape their caging differently around the probe than in the bulk, resulting in the observed acceleration or slowing down.
We interpret that behavior as arising from a modification of  water's structure around the probe.
{\color{black} In agreement with that picture, a previous study\cite{c1} found a decrease of the liquid-liquid transition temperature induced by hydrophobic walls. In our case it will imply an increase of the HDL structure proportion around the probe, leading to an acceleration of the dynamics. Similarly, the decrease of the local density around the hydrophobic probe  also accelerates the dynamics.}
At $\rho=1 g/cm^{3}$ the diffusion of the liquid is slower around the probe than at larger distance, but at lower density  $\rho=0.925 g/cm^{3}$ the probe accelerates the liquid in its surrounding. This result suggests a different modification of the structure near the probe for low and high density.
From these results we expect that the probe induces a looser caging around it at low density while it induces a tighter caging at higher density.
The modification of the diffusion around the probe increases at lower temperatures, a result that will lead to a crossing between the low  and high density diffusion at low temperature. We will find this crossing around $T=215 K$.
Figure \ref{f2} also shows that the modification of the diffusion around the probe is larger at low than at high density, suggesting a larger structural modification at low density or a larger correlation length.
Notice that for large timescales the mean square displacement of the liquid around the probe merge with the bulk mean square displacement, as the water molecules eventually escape from the vicinity of the probe. The characteristic time $\tau$ for that merging increases at low temperature (we find $\tau\approx 2 ns$ for $T=250 K$, $\tau\approx 10 ns$ for $T=230 K$ and $\tau > 40 ns$ for $T=220 K$) due to the relaxation time increase and the increase of the size of the region influenced by the probe. To correct from this effect we will extract in the following, the diffusion coefficient near the probe from intermediate time scales corresponding to $t\approx \tau/10$. 

\begin{figure}
\centering
\includegraphics[height=7.0 cm]{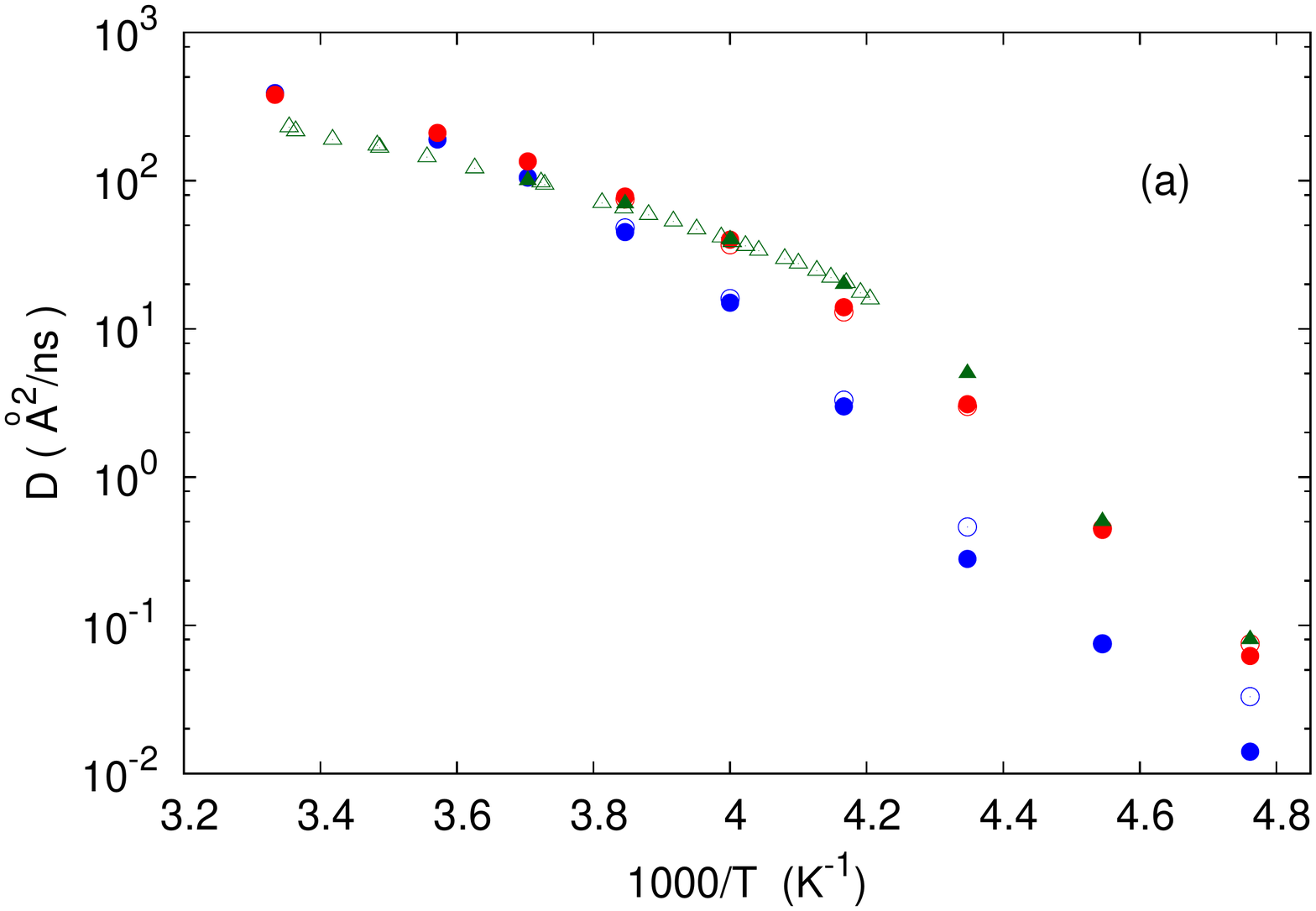}
\includegraphics[height=7.0 cm]{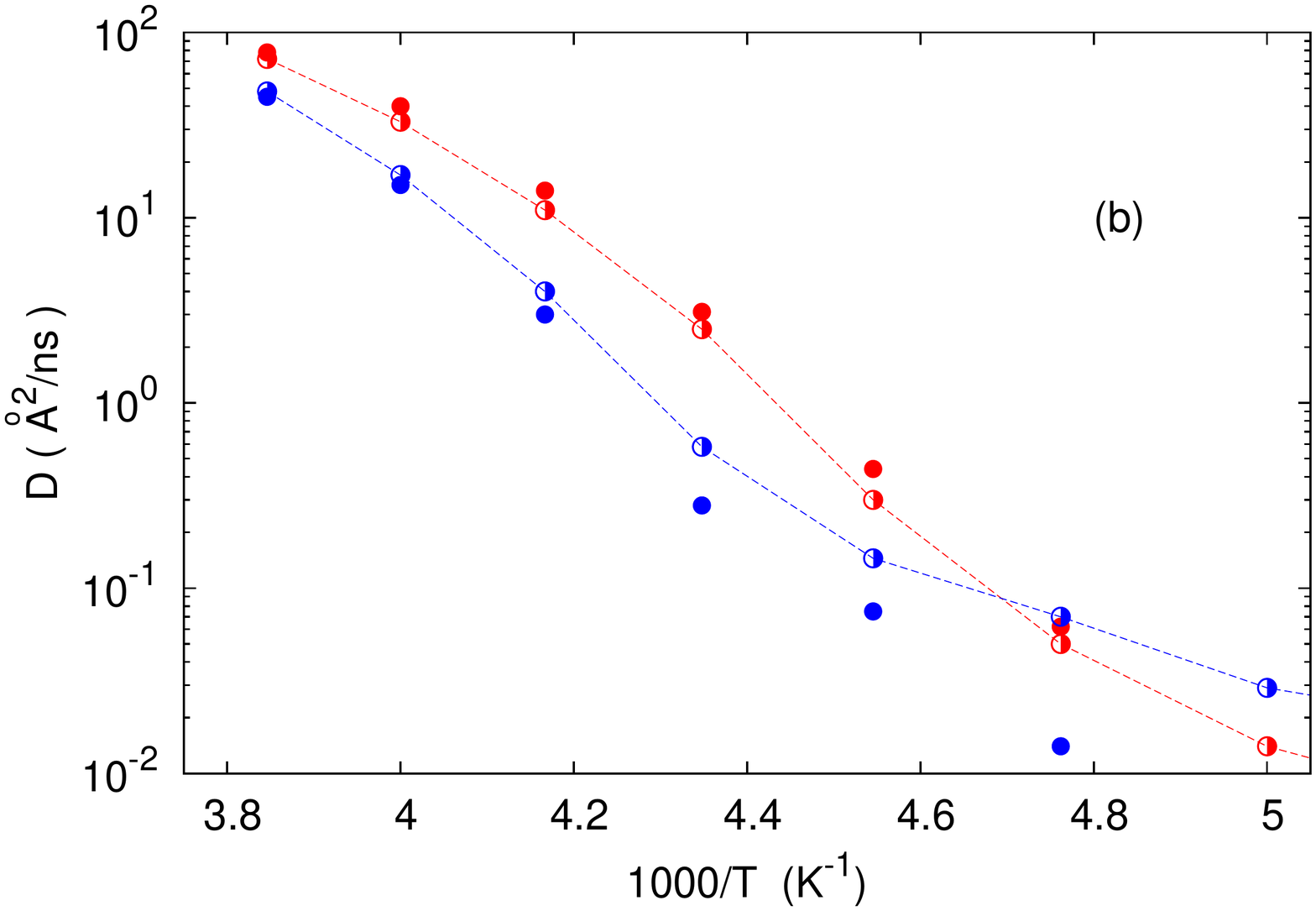}

\caption{(color online) Diffusion coefficient computed versus temperature for two different water densities. The triangles are experimental data: full green triangles \cite{exp},  empty triangles \cite{exp2}.
For both Figures: Red full circles: pure water $\rho=1 g/cm^{3}$; Blue full circles: pure water  $\rho=0.92 g/cm^{3}$.
(a)  Empty circles simulations with the probe inside the box. From top to bottom Red empty circles $\rho=1 g/cm^{3}$ and Blue empty circles $\rho=0.925 g/cm^{3}$. 
(b) Half empty circles: diffusion around the probe ($r<10$\AA\ at $t=0$). From top to bottom: Red half empty circles $\rho=1 g/cm^{3}$ and Blue half empty circles $\rho=0.925 g/cm^{3}$. Lines are guides to the eye.} 
\label{f1}
\end{figure}

Figure \ref{f1} shows the evolution of the diffusion coefficient with temperature at the two densities of the study, in the whole box containing passive probes, 
and in pure water.
The TIP5P simulation model used in this work compares relatively well with the experimental data\cite{exp,exp2} (green and white triangles)  leading to a slightly more viscous liquid (or smaller effective temperatures) than real water.
Interestingly enough we observe in Figure  \ref{f1}a  a divergence between pure water simulations and simulations containing a passive probe at low temperature.
For the later, the curves corresponding to the two densities tend to merge at low temperatures. 
Considering now the diffusion around the probe in Figure \ref{f1}b we observe a crossover around $T=215 K$, below that temperature the diffusion around the probe becomes faster at low than at high density. While at high temperature the water dynamics is the same around the probe than in the bulk, as the temperature decreases the diffusion near the probe becomes faster than in the bulk for $\rho=0.925 g/cm^{3}$ and smaller for $\rho=1 g/cm^{3}$. That tendency increases at low temperature leading eventually to a crossover.

\begin{figure}
\centering
\includegraphics[height=7.0 cm]{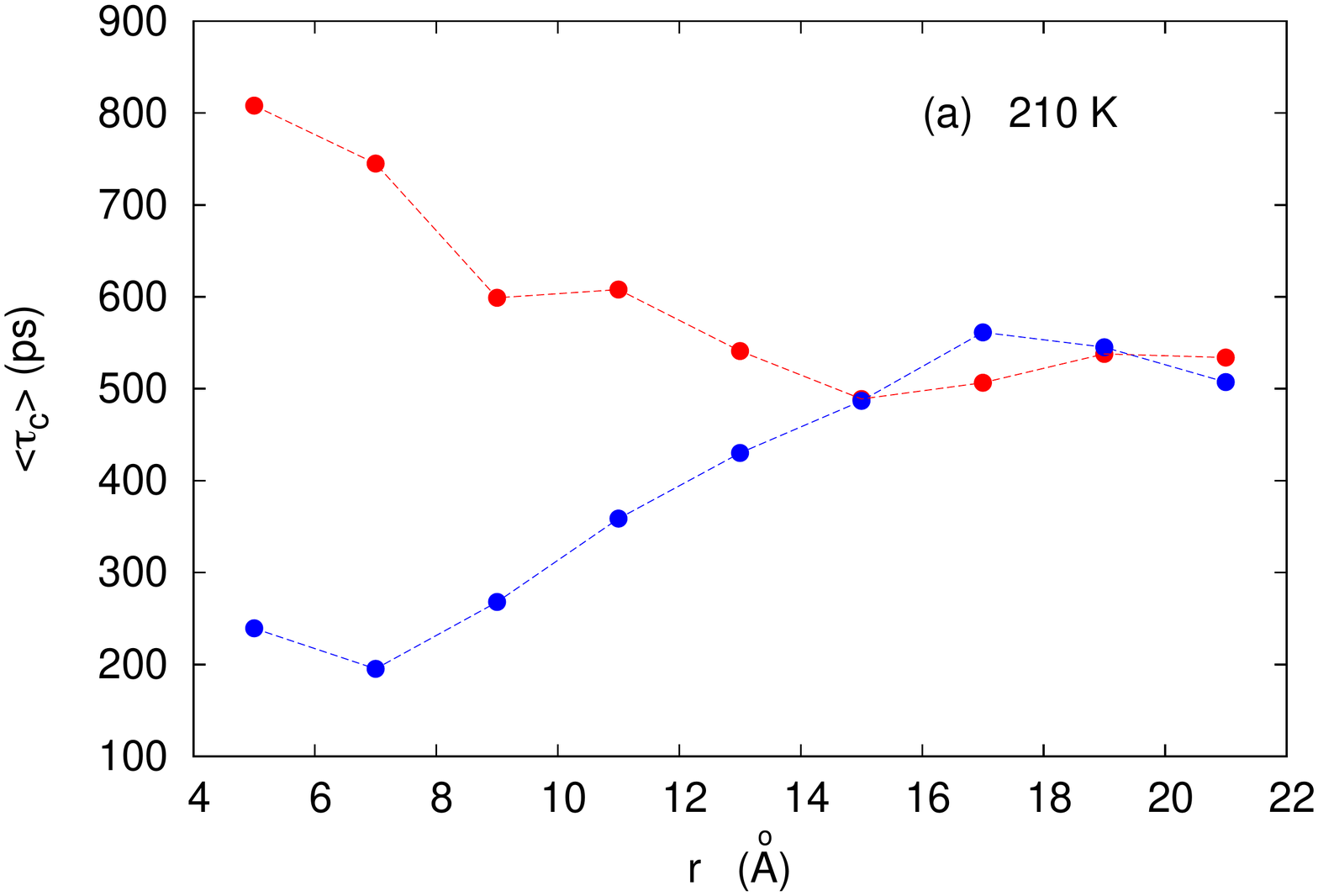}
\includegraphics[height=7.0 cm]{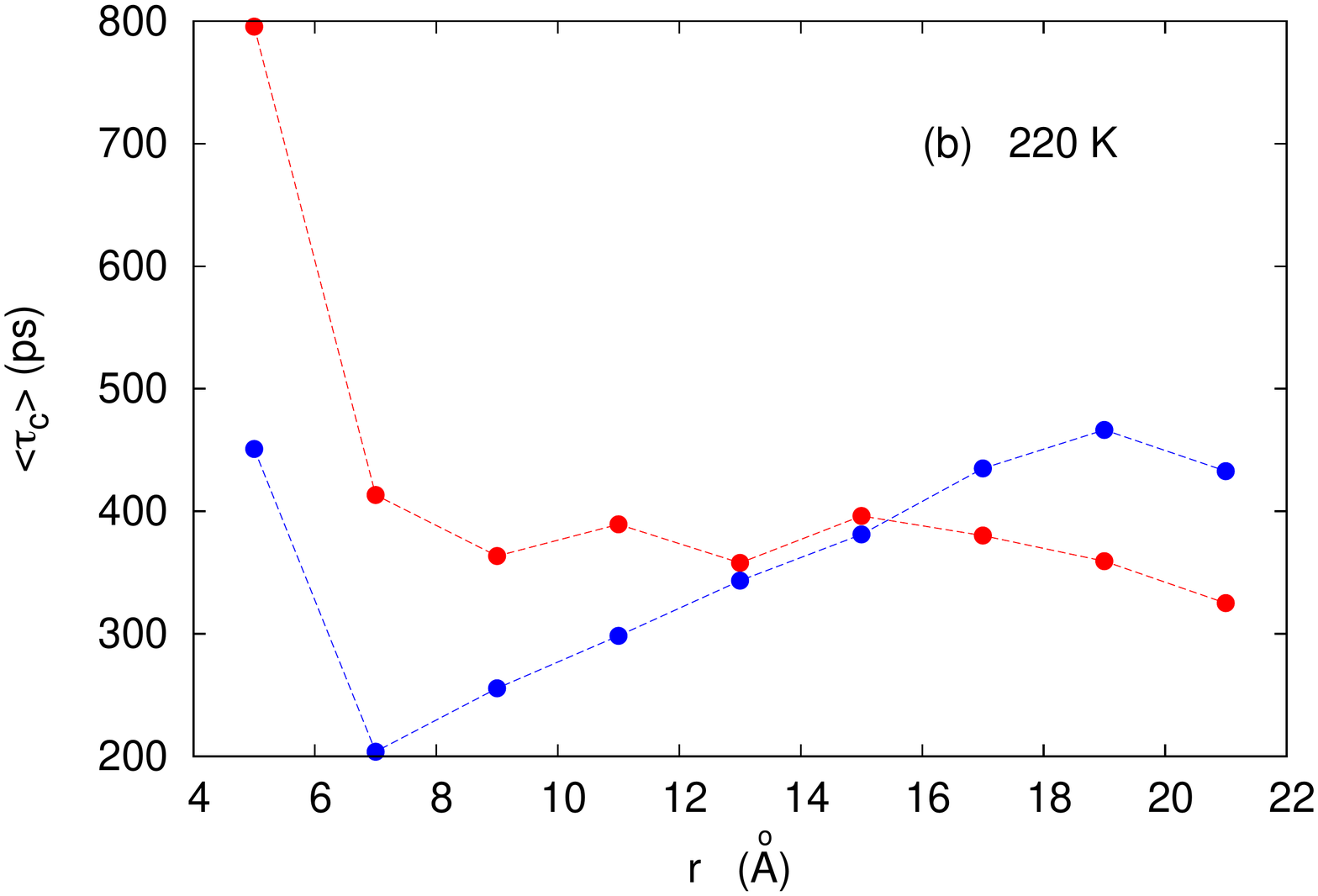}

\includegraphics[height=7.0 cm]{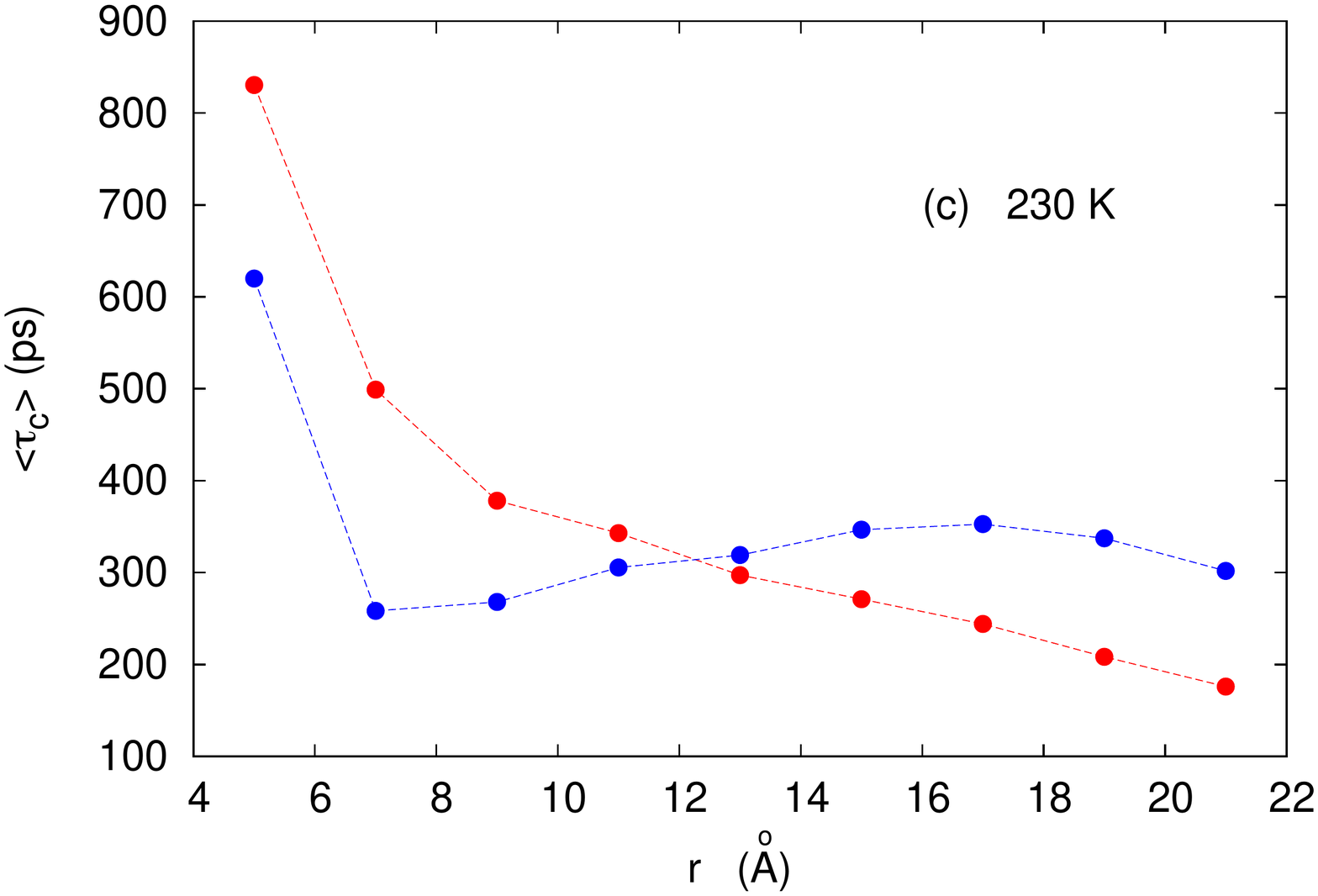}

\caption{\color{black}{(color online) Mean cage survival time $<\tau_{c}>$ for supercooled water at various distances from the motor.
 Red full circles: $\rho=1 g/cm^{3}$; Blue full circles:  $\rho=0.925 g/cm^{3}$.
 (a) T=210 K, (b) T=220 K, (c) T=230K. Lines are guides to the eye.} }
\label{f1v}
\end{figure}

{\color{black} To better understand the cage breaking modification around the motor, Figure \ref{f1v} shows the mean cage survival time of water\cite{dd10} as a function of its distance from the motor. To calculate $<\tau_{c}>$ we define molecules involved in a cage breaking process at a particular time $t_{0}$, as molecules that move at time $t_{0}$ to distance larger than $\Delta r=1 $\AA\ in a time lapse $\Delta t= 1 ps$. Notice that with our definition $<\tau_{c}>$ depends on the choice of the characteristic distance $\Delta r$, however we are only interested in the evolution of the curves with $r$ in our study . We then divide the distance $r$ to the motor by segments of $2$\AA, and calculate $<\tau_{c}>$ for the whole set of molecules inside each shell. Figure \ref{f1v} shows that $<\tau_{c}>$ is larger around the probe for $\rho=1 g/cm^{3}$ and then decreases at larger distances, and we observe the opposite behavior for $\rho=0.925 g/cm^{3}$. 
This leads to a crossing of the two curves at a distance from the probe that increases when the temperature decreases.
This effect results from the increase of  correlation lengths at low temperature, leading to an increase of the domain affected by the motor's presence.
}

If the origin of this dynamical crossover is due to a structural modification around the probe we expect some sort of structural crossover around the probe at the same temperature.
Consequently we will now investigate the structural modifications around the probe and away from it.

\subsubsection{Structural effects}

We have seen in the previous  section that depending on water's density we observe a slow down or an acceleration of the dynamics around the probe.
While dynamical slow down are the most usual, slow downs or accelerations of the dynamics have been reported under confinement in supercooled liquids.
 Previous studies reported that for confinement walls rough enough the dynamics slows down while for smooth walls it  accelerates. For water the hydrophobicity {\color{black}has a similar effect and following previous studies\cite{c1,c2,c3,c4,c5,d1,d2,d3,d4,d5,d6,d7,d8,d9,d10,d11,d12,d13,d14,d15,d16,d17,conf1,conf2,conf3,conf4,conf5,conf6,conf7}  we expect a slowing down for hydrophilic probes and an acceleration for hydrophobic probes as the one we use. For hydrophobic pores the local density of water near the wall decreases, due to the repulsion, explaining the acceleration of the dynamics.}

{\color{black} However, if we find an acceleration of the dynamics around the motor at low density, we also find a slowing down at larger density. A simple explanation for these different behaviors is that they originate from the local structural modification of water around the probe.
We will detail that explanation in two different but related pictures.}
The first is a local modification of the structure of {\color{black}'bulk'} water in the probe's vicinity. 
For example, a more structured water {\color{black} (i.e. an increase of LDL proportion)} promoted by the presence of the probe will  induce a slowing down {\color{black} due to the larger viscosity of LDL}.
The second is the correlation between the probe and the water molecules {\color{black} directly} surrounding it.
For example, if the probe pushes the water molecules surrounding it to large enough distance, it will create {\color{black} a local increase of free volume and }loose caging promoting diffusion.
{\color{black}   The main difference between the two pictures is that in the second picture we consider the first shell of water molecules surrounding the motor, while in the first picture we consider water molecules at slightly larger distance from the probe.}
{\color{black}To test the first picture we will use the radial distribution function between water molecules located around the probe,
while for the second picture we will use the radial distribution function between the probe and the water molecules.}

We show the water structure around the probe in Figure 3, using the radial distribution function (RDF) between water oxygen atoms in a $6$ \AA\ radius around the probe, compared to the same RDF at larger distance from the probe.  
The local structure displayed has been corrected from the exclusion of the volume of the probe\cite{confine} with the simple formula:
\begin{equation}
g(r,r<R,T)=g_{0}(r,r<R,T).K(r)
\end{equation}
The factor $K(r)$ is directly linked to the probe's geometry.
We evaluate that factor, from the hypothesis that at high enough temperature, the water center of masses radial distribution function is only slightly affected by the probe's  presence. 
With this approximation neglecting the probe's effect at high temperature, the factor $K(r)$ can be obtained from a simple ratio of the RDFs at high temperature (here 260K).
\begin{equation}
K(r)={{g_{0}(r,T_{0})} \over {g_{0}(r,r<R,T_{0})}}
\end{equation}


\begin{figure}[H]
\centering
\includegraphics[height=7.0 cm]{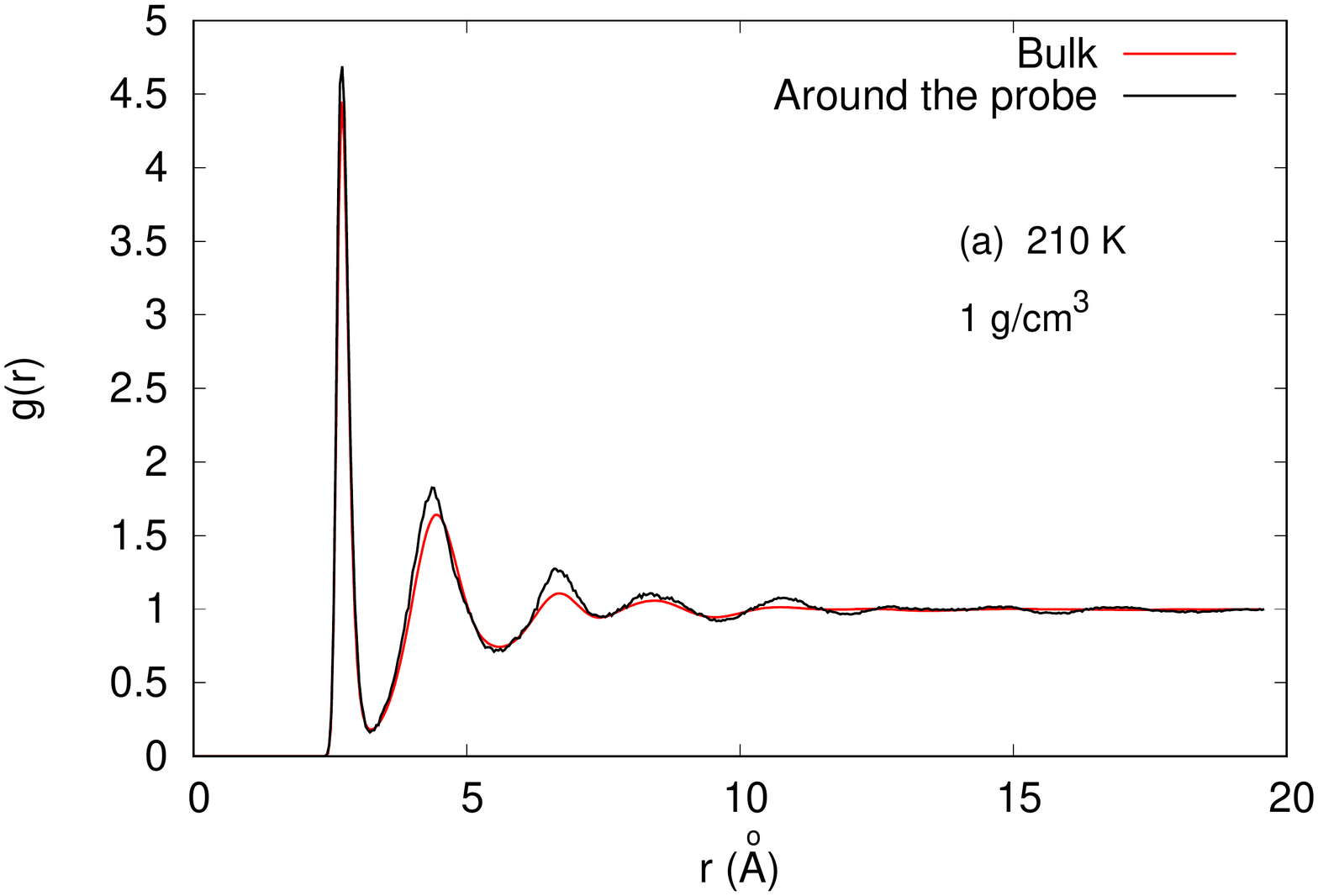}
\includegraphics[height=7.0 cm]{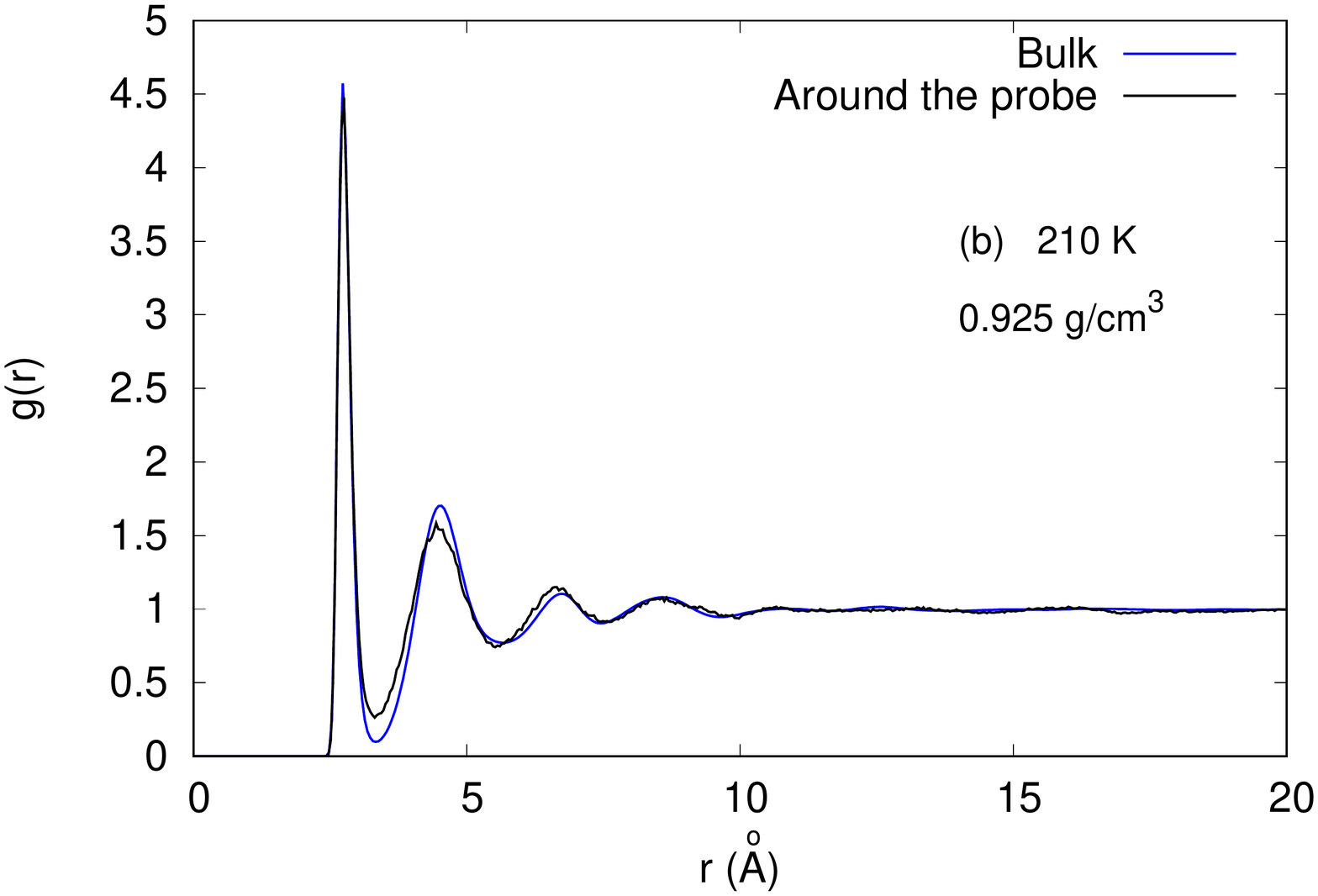}

\caption{(color online) Radial distribution function between oxygen atoms of water molecules g(r) in the bulk compared to the same function with one molecule at a distance $r<6$\AA\ from the motor. The density is: (a) $\rho=1 g/cm^{3}$; (b) $\rho=0.925 g/cm^{3}$.} 
\label{f3}
\end{figure}

Bulk liquid water is more structured at the lower  density  $\rho=0.925g/cm^{3}$  than at $\rho=1g/cm^{3}$ but around the probe the structural difference between the two liquids decreases.
At low density, water near the probe is even less structured than bulk water at high density, and we see a similar effect in Figure \ref{f3} for high density with water near the probe being more structured than low density bulk water.
These modifications of the structure around the probe  result in an increase or a decrease of the diffusion as the cages are less or more compacts, respectively for high and low density.

\begin{figure}[H]
\centering
\includegraphics[height=7.0 cm]{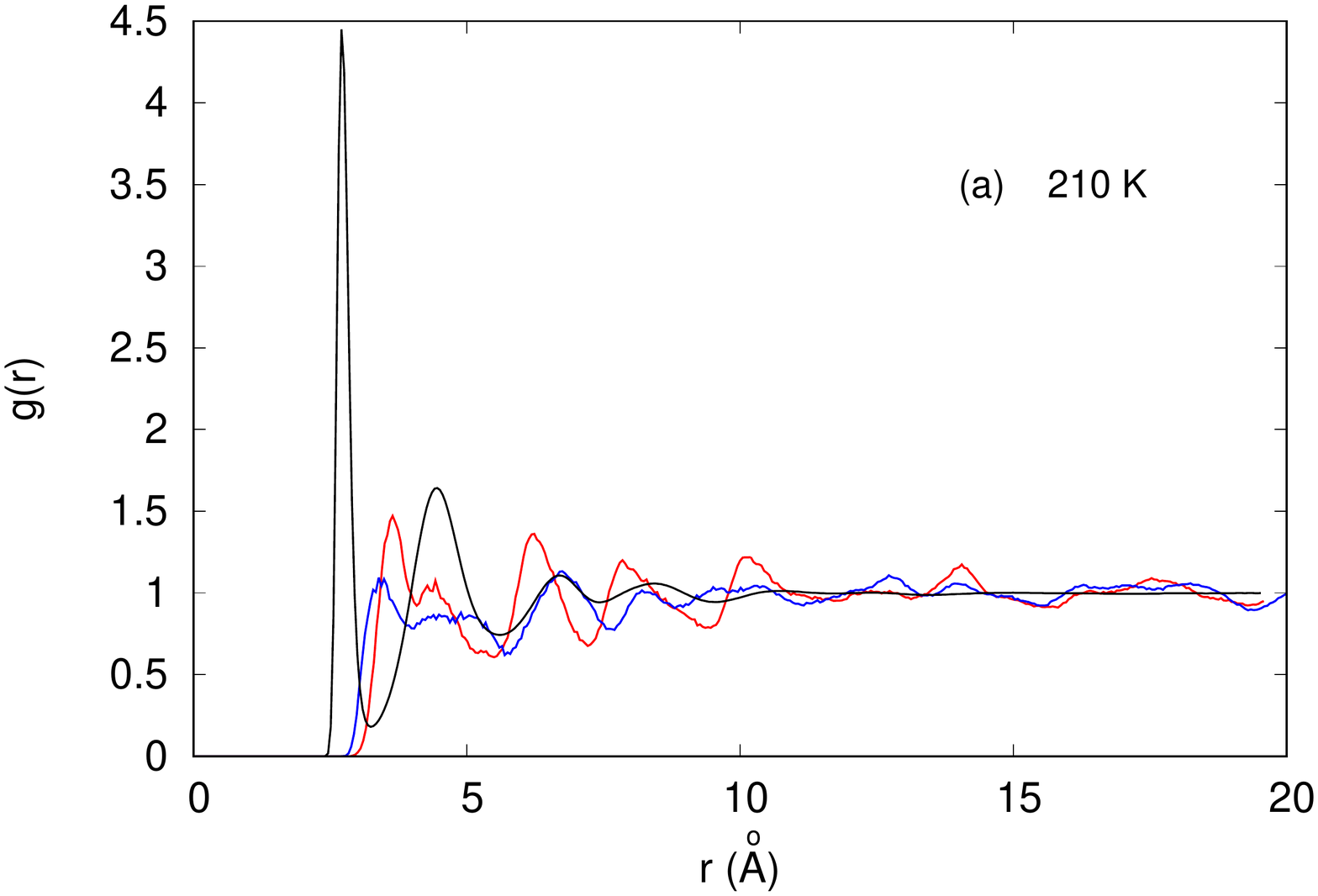}
\includegraphics[height=7.0 cm]{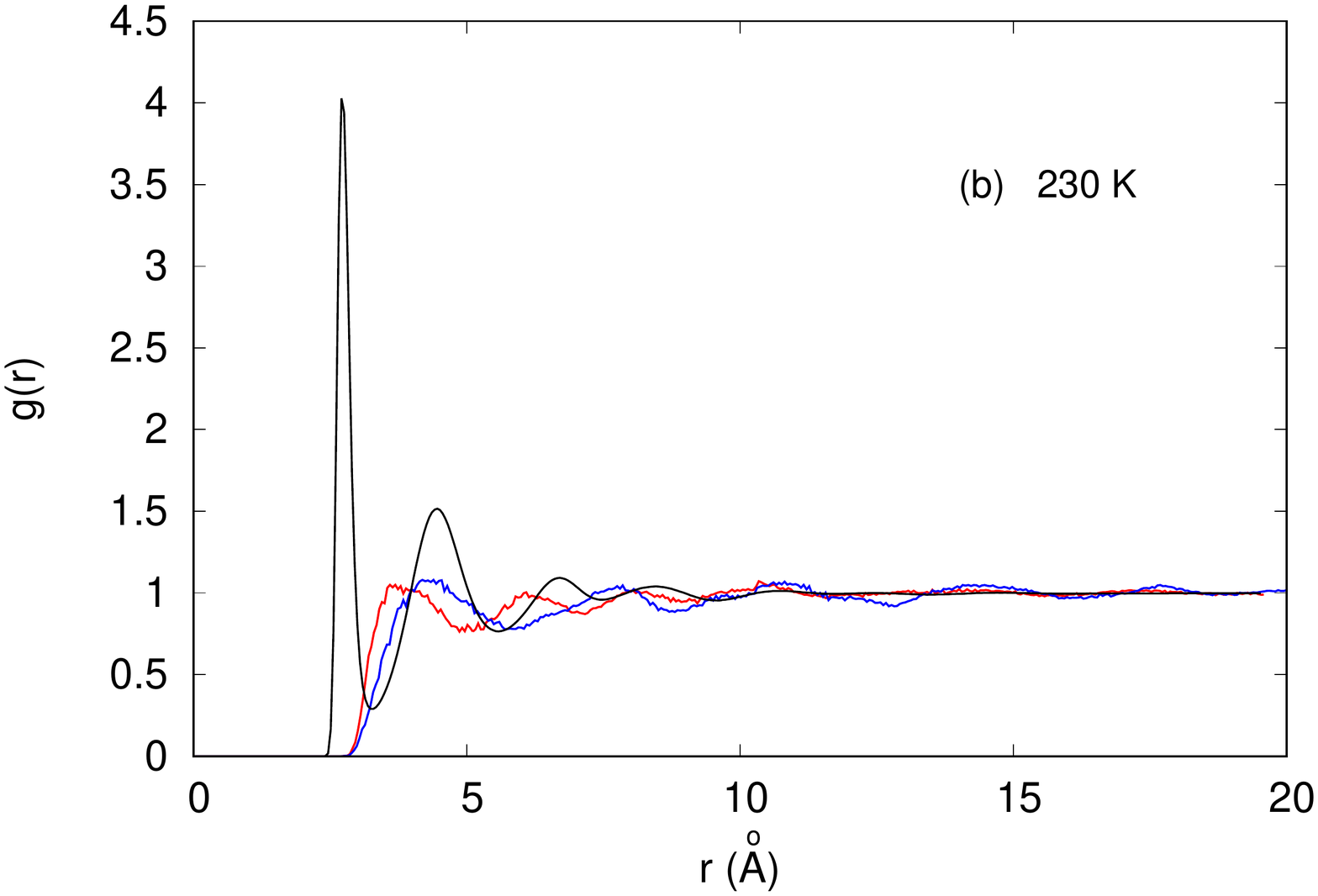}
\caption{(color online) Radial distribution function (RDF) g(r) between the probe and the water centers of masses, compared with the RDF between water molecules at two different temperatures; (a) T=210K, (b) T=230K.
Red curve:$\rho=1 g/cm^{3}$, Blue curve: $\rho=0.925 g/cm^{3}$; Black curve: RDF between CoM of water molecules $\rho=1 g/cm^{3}$.} 
\label{f4}
\end{figure}

The second explanation for the modification of the dynamics comes from the contact of the probe to its  water neighbors.
To test that explanation we display in Figure \ref{f4} the radial distribution function between the probe and the water center of masses at different temperatures (above and below the crossover).
At high density, the water molecules approach the probe in a more tight way and the caging of the water molecules around the probe is larger than at low density. 
When the temperature is lower, the structural modification propagates to larger distances. 
These results agree also well with the observed dynamical modifications induced by the probe.
Consequently the two explanations act in the same direction and add their contributions to the dynamics modifications.

\begin{figure}[H]
\centering
\includegraphics[height=7.0 cm]{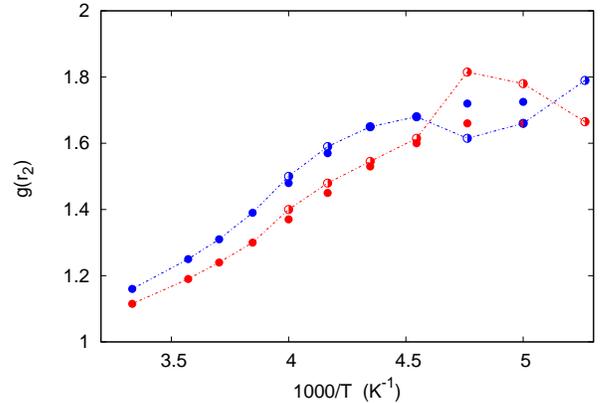}
\caption{(color online) Evolution of the height of the second peak of the oxygen-oxygen radial distribution function with temperature.
Full red circles: Pure water $\rho=1 g/cm^{3}$; Full blue circles: Pure water $\rho=0.92 g/cm^{3}$; Half red circles: around the probe ($r<6$\AA) $\rho=1 g/cm^{3}$; Half blue circles: around the probe ($r<6$\AA) $\rho=0.925 g/cm^{3}$.
$r_{2}=4.5$ \AA.} 
\label{f5}
\end{figure}

In Figure \ref{f5} we display the evolution of the size of the second peak of the oxygen-oxygen radial distribution function with temperature.
The second neighbors peak is located at $r=4.5$\AA\ with a size typically around $1.7$ for LDL water and $1.3$ for HDL water.
Its size in Figure \ref{f5}  permits  to evaluate the proportion of LDL and HDL structures in water at the conditions of study.
 As the temperature decreases, the proportion of LDL structure increases in the bulk, the peak tending to a maximum value of $1.7$ both at high and low densities.
The proportion of LDL in the simulation box for $\rho=1 g/cm^{3}$ and $\rho=0.925 g/cm^{3}$ merge in the Figure for $T\leq 210 K$.
Around the probe however, the evolution of the structure is different.
For $\rho=0.925 g/cm^{3}$ the proportion of LDL increases faster than for the bulk, up to 1.7 for $T=230 K$ and then decreases, while for $\rho=1 g/cm^{3}$ 
the difference is small with the bulk up to $T=215 K$ and then increases to 1.8 for $T\leq 210K$. 
Interestingly enough the two curves cross  at the same temperature than the diffusion coefficients in Figure \ref{f1}. Below that temperature the proportion of LDL water is larger around the probe at high than at low density.
These results suggest that the larger diffusion at low temperature for $\rho=0.925 g/cm^{3}$ arises from a decrease of the proportion of LDL water, leading to the crossover in the diffusion coefficients around the probe observed in Figure \ref{f1}.

\subsection{Active stimuli}

\subsubsection{Dynamical effects}

We will now study the behavior of supercooled water under  active stimuli, that is when the motor is on.
{\color{black}  LDL water will see a faster folding relative to its characteristic relaxation time, because water relaxation time is smaller for the HDL than for the LDL.
From that perspective, even if the folding time is much smaller than the relaxation times, we expect stronger effects in LDL structure\cite{pre}.
}
 When the motor is active, it periodically breaks the water caging surrounding it,  creating excitations inside the medium and inducing diffusion.
 As a result we expect an increase of diffusion whatever the density studied.

\begin{figure}
	\centering
	\includegraphics[height=7.0 cm]{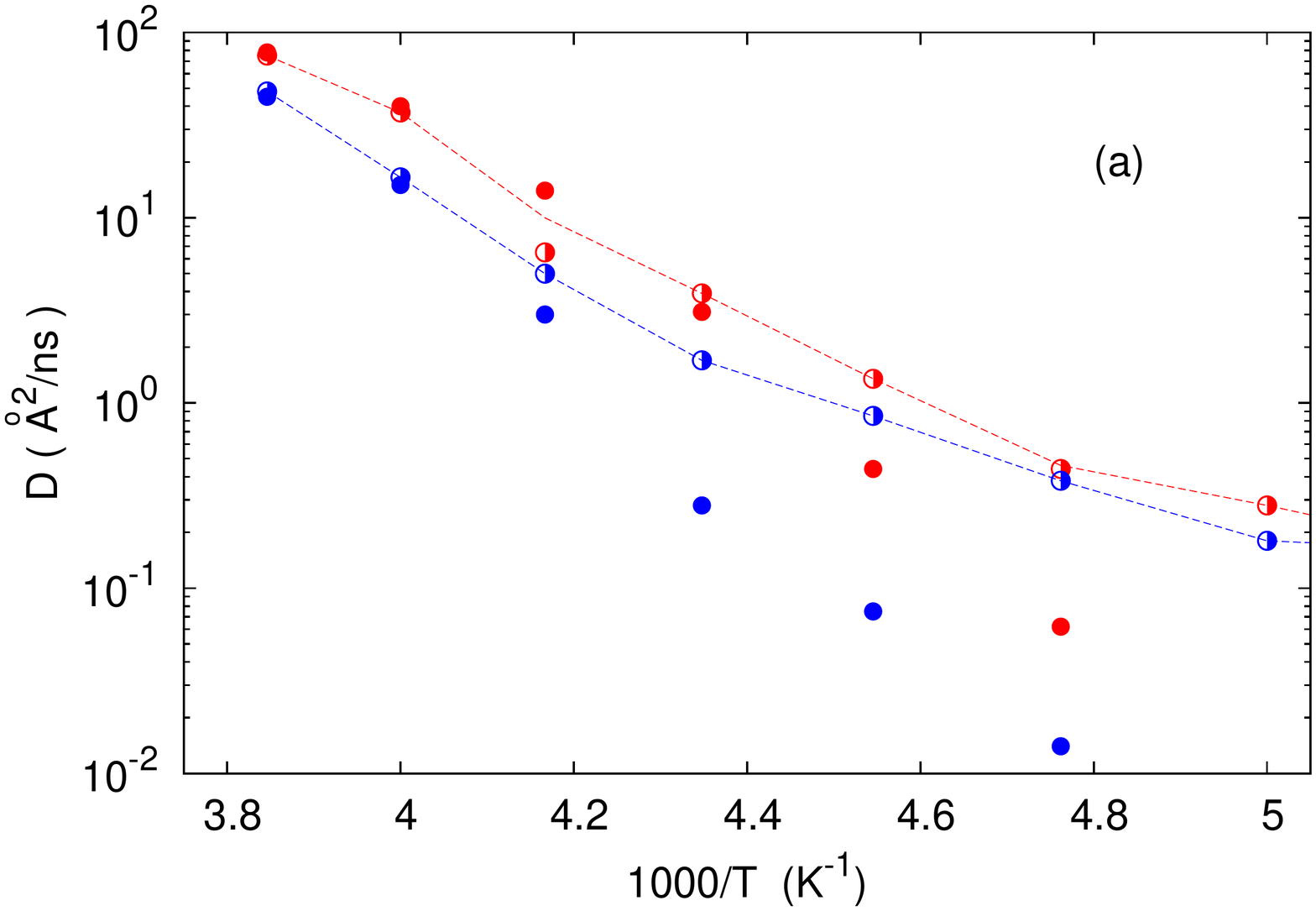}
	\includegraphics[height=7.0 cm]{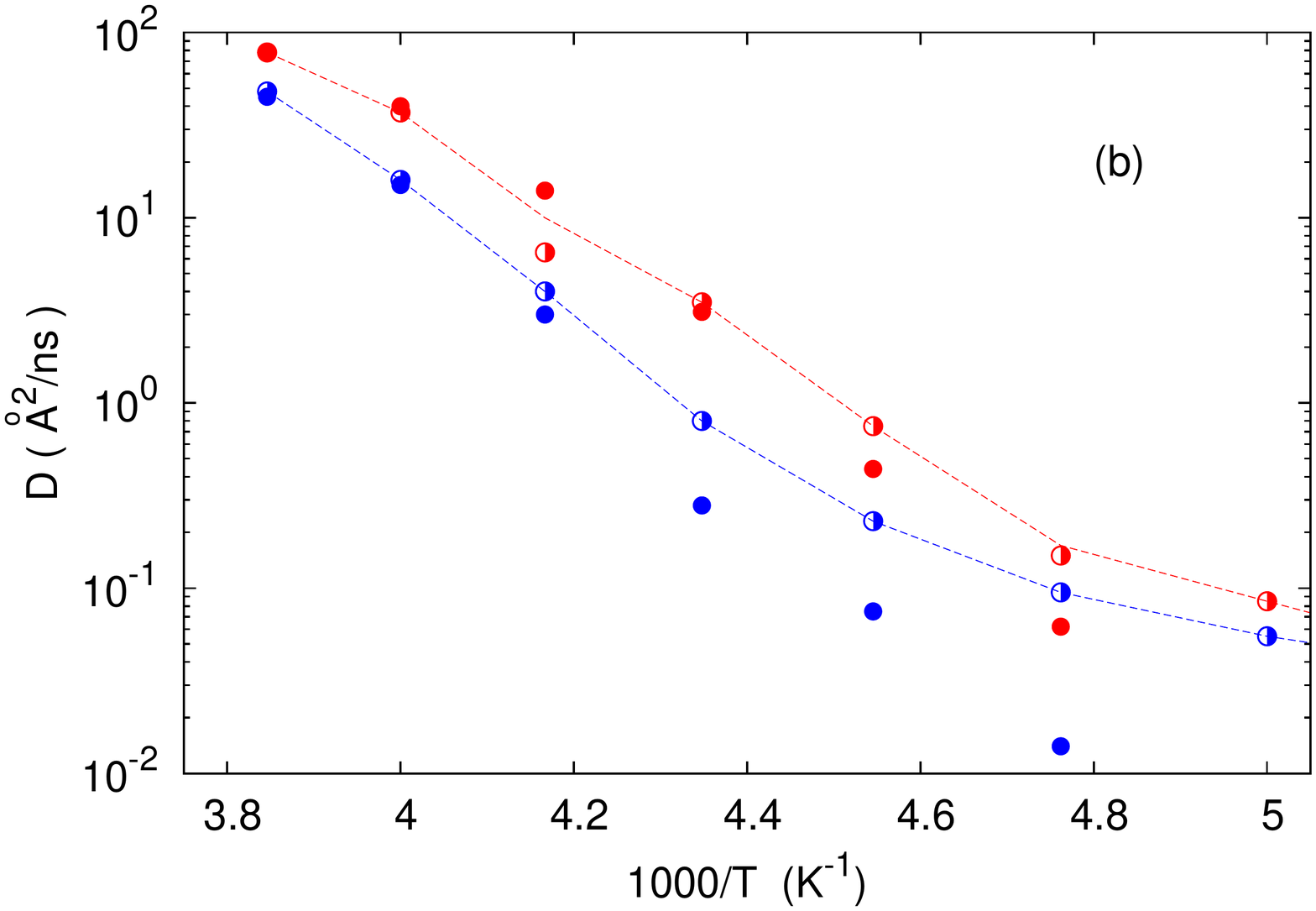}
	
	\caption{(color online) Diffusion coefficient when the motor is active (on: Half empty circles) or not (off: Full circles).
	Red circles: $\rho=1 g/cm^{3}$; Blue circles: $\rho=0.925 g/cm^{3}$; 
		(a) Half empty circles: diffusion around the probe ($r<10$ \AA); (b) Half empty circles: diffusion in the whole box.
	} 
	\label{f6}
\end{figure}

 That's what we see in Figures \ref{f6}, however the diffusion increase is not the same at the two densities.
 It is larger at low density and that behavior increases when the temperature drops.
 As a result the high and low density diffusions tend to merge at low temperature.
If we do not find a crossing of the high and low density curves in our data, it is possible that a crossing appears  at lower temperatures.

The difference between water  diffusion subjected to an active stimulus and  bulk water diffusion increases when the temperature decreases.
That behavior can be rationalized if the diffusion induced by the active stimulus doesn't depend on the temperature or depends weakly on it.
In that picture, we have:
\begin{equation}
D_{on}(T)=D_{off}(T) + D_{induced}(\nu_{stimuli},T) \label{eq1}
\end{equation}
Where $D_{induced}(\nu_{stimuli},T)$ is the diffusion induced by the stimuli, and $\nu_{stimuli}$ is the stimuli frequency.
As we expect the second term in equation \ref{eq1} to depend only weakly on the temperature and $D_{off}$ tends to zero at low temperature, $D_{on}$ will tend in that picture to the second term at low enough temperatures.
Interestingly, Figure \ref{f6} suggests that  $D_{induced}(\nu_{stimuli},T)$ is approximately the same for both densities, as the curves tend to merge at low temperature.
Note that from the difference $D_{on}(T)-D_{off}(T)$ we found that $D_{induced}(\nu_{stimuli},T)$ actually depends weakly on $T$, decreasing slowly when $T$ decreases.

\begin{figure}
	\centering
	\includegraphics[height=7.0 cm]{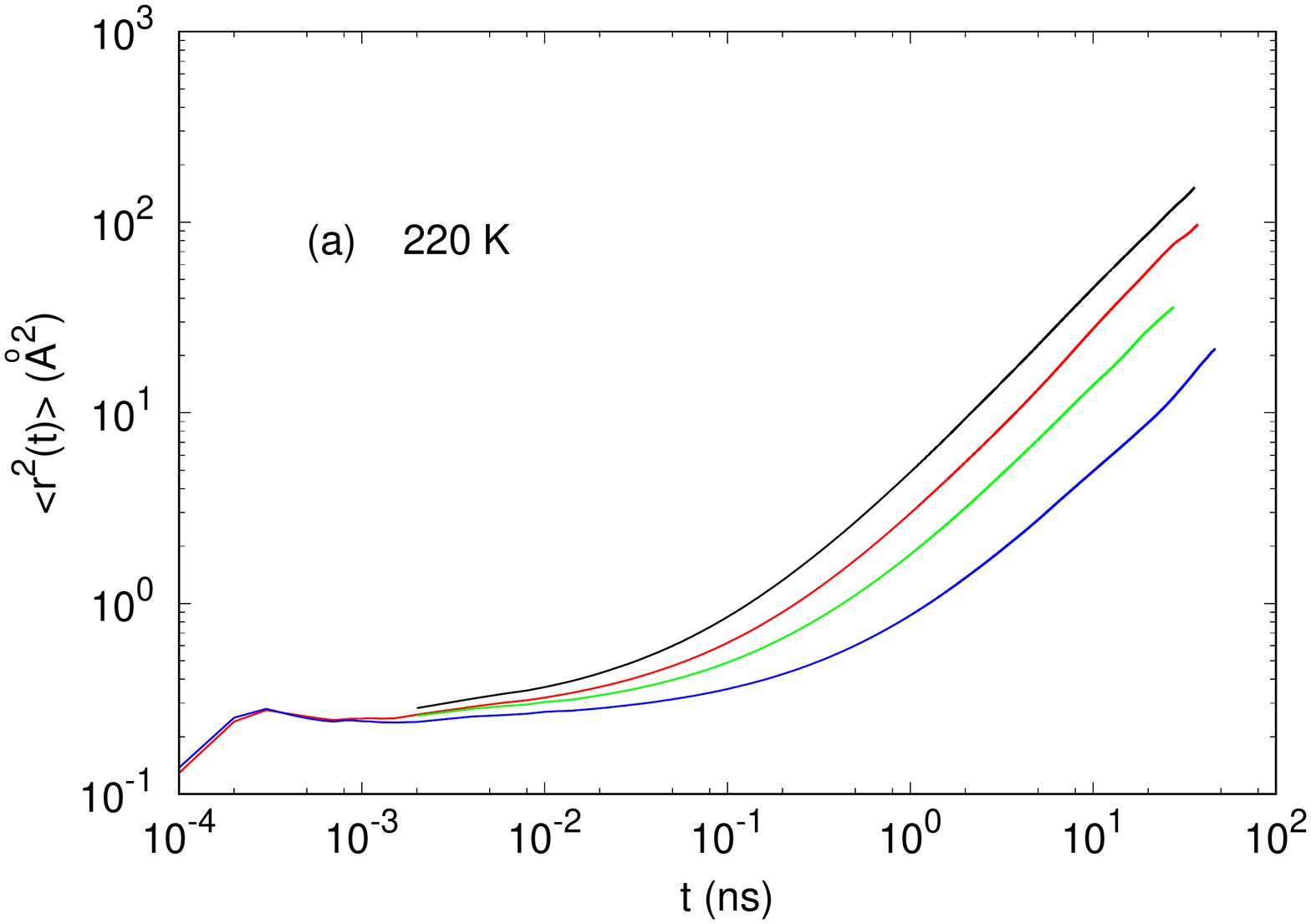}
	\includegraphics[height=7.0 cm]{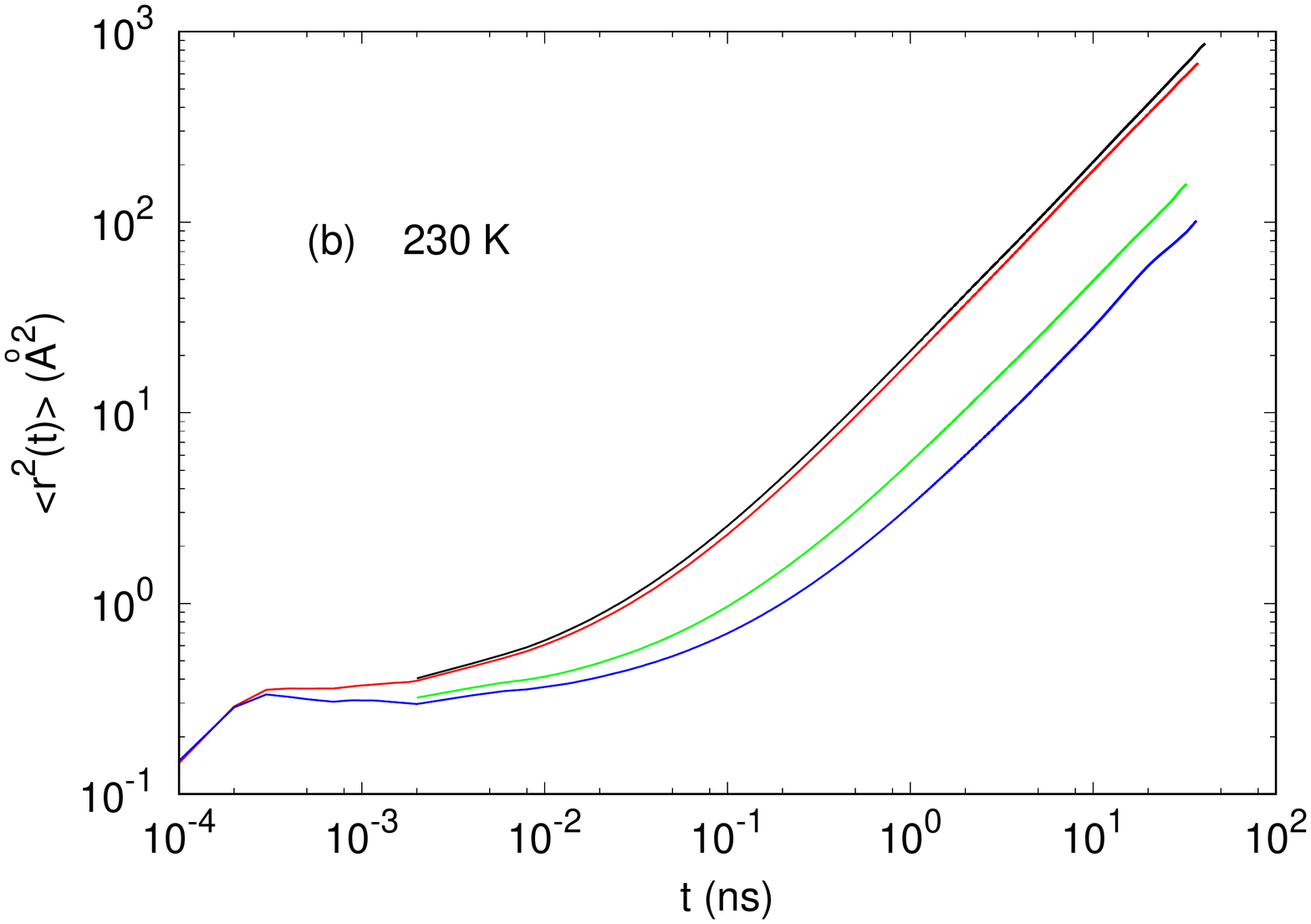}
	\includegraphics[height=7.0 cm]{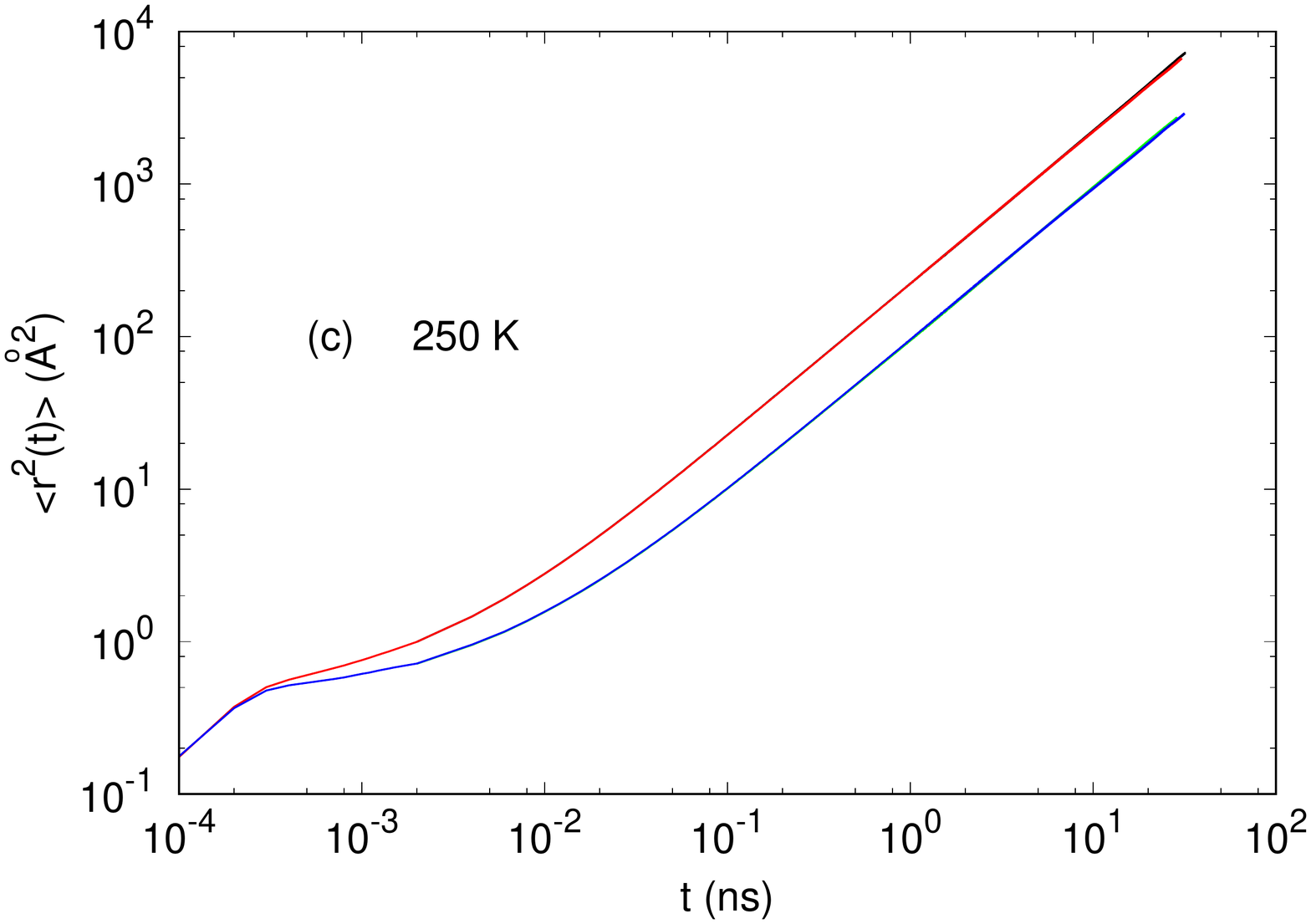}
	
\caption{(color online) Comparison of the mean square displacement of the center of masses of water molecules when the motor is on or off, for two different densities.
The temperature is (a) $T=220 K$, (b) $T=230 K$, (c) $T=250 K$.
From top to bottom in each Figure: 
Black line: water MSD around the probe ($r<10$\AA\ at $t=0$) for $\rho=1 g/cm^{3}$. 
Red line: bulk water MSD for $\rho=1 g/cm^{3}$. 
Green line: water MSD around the probe ($r<10$\AA\ at $t=0$) for $\rho=0.925 g/cm^{3}$. 
and Blue line: bulk water MSD for $\rho=0.925 g/cm^{3}$.} 
	\label{f7}
\end{figure}

To improve our understanding on the diffusion behavior, in Figure \ref{f7} we display the water mean square displacement at different temperatures and densities when the motor is active or not.
The Figure shows that the displacement enhancement induced by the motor is larger at low density and also increases when the temperature drops. 
The enhancement also appears at higher temperature at low density ($T=230K$) than at high density ($T=220K$).
These effects appear in the plateau regime, the cage escaping process being faster when water is subject to the motor's stimuli.
The stimuli induced cage breaking model\cite{cage}, that is the hypothesis that the unfolding motor breaks water cages around it, permitting elementary diffusive processes, explains that observation.
Notice also that these behaviors are remarkably similar to what we observed when the motor was passive, acting as a simple probe. 
Finally, the difference between the two densities suggests different liquid properties. In an other picture it suggests  that the cage breaking effect is larger or easier at low density.

\subsubsection{Structural effects}

\begin{figure}[H]
	\centering
	\includegraphics[height=7.0 cm]{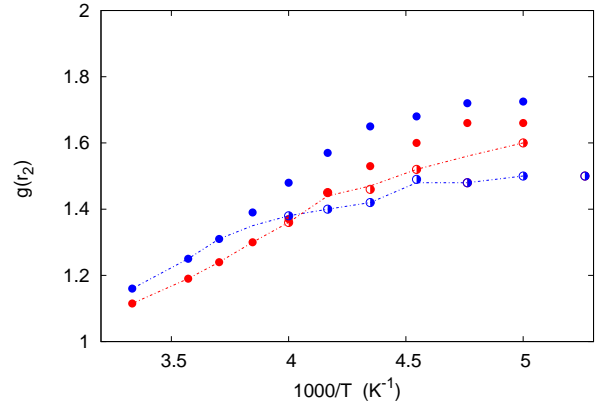}
		
	\caption{(color online) Evolution of the height of the second peak of the oxygen-oxygen radial distribution function with temperature.  $r_{2}=4.5$ \AA. 
Full red circles: Pure water $\rho=1 g/cm^{3}$; Full blue circles: Pure water $\rho=0.92 g/cm^{3}$; Half red circles: around the active motor ($r<6$\AA) $\rho=1 g/cm^{3}$; Half blue circles: around the active motor ($r<6$\AA) $\rho=0.925 g/cm^{3}$.
	} 
	\label{f8}
\end{figure}

In Figure \ref{f8} we show the evolution of the second peak of the oxygen-oxygen RDF for an active stimulus.
That Figure can be compared with Figure \ref{f5} that is the same Figure for a passive stimulus.
In contrast to what we observed for a passive stimulus in Figure \ref{f5}, Figure \ref{f8} shows that water submitted to an active stimulus is less structured than bulk water at low temperature.
This effect appears around $250K$ at low density and around $230K$ at high density.
We conclude that our motor when active favors HDL water, while it favors LDL water when acting as a passive probe.
As for the passive stimulus, we observe a crossing of the curves, here around $245 K$,  water becoming more structured at high than at low density for low temperatures.
However the temperature range for which high density water is more structured is larger here than for the passive stimulus.
Note that a second crossing  appears  at lower temperature around $190K$ as for the passive stimulus, however we cannot be conclusive about that point due to the  uncertainty of our results at that temperature.
\vskip 1cm
\section{Conclusion}

In this paper we studied the behavior of supercooled water  subject to different stimuli from an active or passive diluted azobenzene hydrophobic probe.
When the azobenzene molecular motor doesn't fold, it acts as a passive probe, modifying the structure of water around it, while when the motor is active, it induces elementary diffusion processes inside the medium acting mainly on the dynamics. 
We found that the passive probe induces ever an acceleration or a slowing down of the diffusion process around it depending on the density of water, while the active probe induces acceleration only. 
We found a crossover between the diffusion coefficients for the two densities near the passive probe, around T = 215K. The dynamical crossover is associated to a modification of the structure of water near the probe. We found a crossover of the proportion of LDL water around the same temperature suggesting that it induces the observed dynamical crossover. The active stimuli increase diffusion for both densities and decrease the proportion of LDL water at low temperature. However we also found for the active stimuli a crossover of the LDL proportion between the two densities of study, showing remarkable similarities between active and passive stimuli results.

\end{document}